\documentclass[structabstract]{aa}  
\usepackage{graphicx}
\usepackage{txfonts}
\usepackage{natbib}
\usepackage{longtable}
\begin{document}
   \title{Primordial star formation: relative impact of H$_2$ three-body rates and initial conditions}


   \author{S. Bovino 
          \inst{1}
         \fnmsep\thanks{Corresponding author email: sbovino@astro.physik.uni-goettingen.de},
          D. R. G. Schleicher
          \inst{1},
         \and
          T. Grassi
          \inst{2}
          }

 \institute{$^1$Institut f\"ur Astrophysik, Georg-August Universit\"at, Friedrich-Hund-Platz 1, 37073, G\"ottingen\\
 $^2$Department of Chemistry, "Sapienza" University of Rome
              P.le A. Moro 5, 00185 Rome\\
             }

   \date{Received ; accepted}

 
  \abstract
   {Population III stars are the first stars in the Universe to form at $z=20-30$ out of a pure hydrogen and helium gas in minihalos of $10^5-10^6$~M$_\odot$ . Cooling and fragmentation is thus regulated via molecular hydrogen. At densities above $10^8$~cm$^{-3}$, the three-body H$_2$ formation rates are particularly important for making the gas fully molecular. These rates were considered to be uncertain by at least a few orders of magnitude.}
   {We explore the impact of new recently derived accurate three-body H$_2$ formation for three different minihalos, and compare them with the results obtained with three-body rates employed in previous other studies.}
   {The calculations were performed with the cosmological hydrodynamics code \textsc{enzo} (release 2.2) coupled with the  chemistry package \textsc{krome} (including a network for primordial chemistry), which was previously shown to be accurate in high-resolution simulations. }
   {While the new rates can shift the point where the gas becomes fully molecular, leading to a different thermal evolution, there is no trivial trend in the way this occurs. While one might naively expect the results to follow the rate coefficients trend, the behavior can vary depending on the dark-matter halo that is explored.}
   {We conclude that employing the correct three-body rates is about equally important as the use of appropriate initial conditions, and that the resulting thermal evolution needs to be calculated for every halo individually.}

   \keywords{Astrochemistry --
                Molecular processes -- ISM: molecules --
                Methods: numerical -- evolution - cosmology: theory - Population III
               }

   \titlerunning{H$_2$ formation rates and initial conditions effect on primordial minihalos}
   \maketitle
%

\section{Introduction}
Chemistry and cooling are known to be very important in regulating gravitational collapse and the fragmentation of the gas \citep{LiKlessenMacLow03, Peters12}.
Since the pioneering work by \citet{Saslaw1967}, it is well-known that molecular hydrogen forms already at $z\sim300$ in the early Universe. Cooling via H$_2$ then allows gravitational collapse and star formation in minihalos with 10$^5$-10$^6$~M$_\odot$ at $z = 20-30$. These Population III stars were thought to be very massive and short lived \citep{Abel2002, Bromm04, Yoshida08}, and subsequently shape their environment via chemical, mechanical, and radiative feedback \citep{Ciardi05, Tornatore07, Schleicher08, Schneider08}. In this way, they provide the initial conditions for subsequent star formation.

Recent studies  suggested that fragmentation occurs, which leads to the formation of clusters and binary systems \citep{TurkScience09, ClarkGlover2011, Greif2011ApJ, Smith11, Greif2012}, while radiative feedback provides upper mass-limits of the order of $\sim50-100$~M$_\odot$ \citep{Hosokawa11,Susa2013}. In particular for the potential formation of low-mass stars, the details of the thermal evolution can be highly relevant, because they influence the resulting fragmentation process.  It is therefore very important to clearly assess the accuracy of the microphysical processes that involve the main coolant, which is molecular hydrogen. 

At low densities H$_2$ begins to form via the H$^-$ path \citep{Saslaw1967,Peebles1968} and becomes an effective coolant around 10~cm$^{-3}$. At high densities, the importance of the three-body (3B) rates increases significantly \citep{Palla1983}, because collisions then scale with the density to the third power. The dominant reaction is given as
\begin{equation}\label{eq:formation}
	\rm H + H + H \rightarrow H_2 + H
\end{equation}
 and its inverse reaction, the collisionally induced dissociation (CID),
 \begin{equation}\label{eq:destruction}
 	\rm H_2 + H \rightarrow H  + H + H
 \end{equation}
comes into play. The three-body H$_2$ formation can also substantially contribute to the heating of the gas, because the binding energy of H$_2$ is released into the gas. 

\begin{figure}
   \centering
   \includegraphics[width=0.45\textwidth]{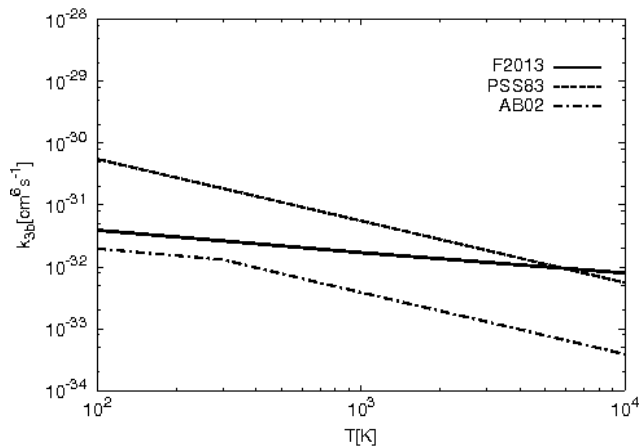}
      \caption{Three-body H$_2$ formation rates as function of the temperature: AB02 from \citet{Abel2002}, F2013 from \citet{Forrey2013}, and PSS83 from \citet{Palla1983}.}
         \label{fig:rates}
   \end{figure}

The uncertainty of the reaction in Eq. \ref{eq:formation} formerly encompassed about two orders of magnitude (see Fig. \ref{fig:rates}). The choice of the rate may accordingly strongly affect the features of the collapsing cloud and its final mass \citep{Glover2008}. The main uncertainty has been explored by \citet{Turk2011ApJ} using adaptive-mesh refinement (AMR) simulations with \textsc{enzo} \citep{Enzo2013} and smoothed particle hydrodynamics (SPH) simulations with Gadget \citep{Springel05}. These authors concluded that the inner region of the collapsing cloud is strongly affected by the rate at which H$_2$ forms, producing lower or higher accretion rates onto the disk. The latter is potentially relevant for disk stability and fragmentation. It also appeared that the different initial conditions employed with the two codes may have a strong impact on the results. While \citet{Turk2011ApJ} focused more on possible discrepancies caused by the use of two different approaches (AMR and SPH), we decided to explore the effect of new 3B rates by employing different initial conditions within the same AMR code. This is more consistent and allows us to assess some additional effects because of the different minihalo selection.

We addressed this uncertainty here by adopting a new accurate three-body H$_2$ formation rate that was recently derived by \citet{Forrey2013}. We calculated the resulting thermal and chemical evolution for three minihalos and compare them with the results obtained from the rates that were previously employed in the literature. A detailed discussion of the new and previously employed rates is given in the following section. The numerical approach employed here is then described in Section \ref{sec:numerics}, including the methodology and the initial setup. Our main results are given in Section \ref{sec:results}, and Section \ref{sect:conclusions} summarizes our main conclusions.

\begin{table*}
	\caption{Three-body formation and destruction rates for the different tests (see text for details).}
	\begin{center}
		\begin{tabular}{ll}
			\hline
			\hline
			& H$_2$ formation rates (cm$^6$s$^{-1}$)\\ 
			\hline
			\hline
			AB02 & 1.3$\times$10$^{-32}(T/300)^{-0.38}$\hspace{0.2cm} ($T \le 300 K$)\\
			AB02 & 1.3$\times$10$^{-32}(T/300)^{-1.00}$\hspace{0.2cm} ($T > 300 K$)\\
			PSS83 &  5.5$\times$10$^{-29}/T$\\
			F2013 & $6.0\times10^{-32}T^{-0.25} + 2.0\times10^{-31}T^{-0.5}$\\
			\hline
			\hline
			& H$_2$ destruction rates (cm$^3$s$^{-1}$)\\
			\hline\hline
			AB02 &  ($1.0670825\times10^{-10}\times T_e^{2.012})/(\exp(4.463/T_e)\times(1 + 0.2472 T_e)^{3.512})$\\
			PSS83 &  5.24$\times 10^{-7}\times T^{-0.485}\exp(-52000/T)$ \\
			F2013 & same as AB02\\
			\hline
		\end{tabular}
	\end{center}
	\label{tab:three-body}
\end{table*}

\section{Three-body H$_2$ formation and CID rates}\label{sec:three-body}
The rate for the CID reaction in Eq. \ref{eq:destruction}, which is the inverse process of Eq. \ref{eq:formation}, was first measured by \citet{Jacobs1967} with experimental studies on shock waves in a  temperature range from 2900~K to 4700~K. The analytic expression for the dissociation rate constants was based on \emph{JANAF}\footnote{http://kinetics.nist.gov/janaf/} equilibrium data and the error was estimated to be about a factor of 2.  However, the comparison with previous experimental data shows differences of an order of magnitude. The \citet{Jacobs1967} rates have been widely used over the years as a starting point to evaluate 3B H$_2$ formation rate (Eq. \ref{eq:formation}) and as benchmark for additional theoretical studies. For instance, \citet{Palla1983} and \citet{Flower2007} reported H$_2$ formation rate coefficients obtained from the measurements of \citet{Jacobs1967} by applying the principle of detailed balance. 
 
 Direct 3B H$_2$ formation rates have been obtained by \citet{Shui1973} by employing phase-space theory (PST) for temperatures ranging from 300~K to 10000~K. A detailed comparison between experimental and theoretical data has been reported and discussed. While the CID rates agreed well with previous existing data \citep{Jacobs1967,Gardiner,Hurle}, no good match was found with the experimentally inferred values for reaction of Eq. \ref{eq:formation}. All the previous studies are based on an indirect evaluation of the 3B formation rates by using data from direct dissociation shock-tube experiments. The only direct measurements reported by \citet{Bennett} exhibited a maximum at a temperature below 3000 K and was three times lower than the theoretical values reported by \citet{Shui1973}. The work by \citet{Bennett} unfortunately has never been considered in astrophysical studies, which always referred to the original \citet{Jacobs1967} work.  The resonance complex theory was employed by \citet{Orel1987} and new rates were provided at temperatures of a few hundred kelvins and were compared with the earlier work by \citet{Whitlock} (who applied a similar method), showing a discrepancy of a factor of 2. Finally, \citet{Esposito2009} reported rate coefficients based on quasi-classical approaches.
 
We can therefore conclude that based on the publications during the past 50 years, it is easy to find large discrepancies between theoretical and experimental data. 
There are at least three different rates that are commonly used in cosmological simulations: (i) the one proposed by \citet{Palla1983} based on the experiment of \citet{Jacobs1967}, which has been obtained by applying the principle of detailed balance to reaction in  Eq. \ref{eq:destruction}; (ii) the extrapolation by \citet{Abel2002} based on a direct calculation of the formation path by \citet{Orel1987} between 10-300 K: this rate shows a sudden change of slope at 300 K that is artificial because of the extrapolation adopted by the authors to higher temperatures; (iii) the \citet{Flower2007} rates, which are based on the work of \citet{Jacobs1967} but were obtained by using a different set of partition functions, and (iv) a new fit used by \citet{ClarkGlover2011} based on the CID rate constants of \citet{Martin1996}. The above rates have been used in several studies and the differences between them have been discussed by \citet{Turk2011ApJ} and \citet{Glover2008}. 

The recent calculation performed by \citet{Forrey2013} is, to our knowledge, the first accurate quantum-mechanical study for the 3B formation of molecular hydrogen. It was obtained by applying the Sturmian theory \citep{Forrey2013a} under the energy sudden approximation \citep{Kramer1964,Khare1978}. These results agree well with the experiment by \citet{Jacobs1967} and with the theoretical results from the PST \citep{Shui1973} and QCT \citep{Orel1987} calculations. It is worth noting that this is the first direct calculation of the formation reaction that is valid for a wide range of temperatures for which no extrapolation has been applied. 

We note that \citet{Forrey2013} also reported an error in the rate calculation by \citet{Flower2007} that was previously pointed out by \citet{Esposito2009} and discussed in \citet{Galli2013}. Because the H$_2$ partition function was  evaluated neglecting the electron spin degeneracy, the final rate was higher by a factor of 4 than that reported by \citet{Palla1983} based on the same CID measurements \citep{Jacobs1967}. The calculation by \citet{Flower2007} thus leads to the same rates as that of \citet{Palla1983} if the correct normalization is employed. Even though the rates by \citet{Flower2007} have been employed in previous studies, here we decided to not include them. From a more physical point of view, one should include the correction pointed out by \citet{Forrey2013}, which leads to a rate equivalent to the \citet{Palla1983} result.

\section{Numerical method}\label{sec:numerics}
\subsection{Cosmological setup}

We performed our simulations by employing the cosmological hydrodynamics AMR code \textsc{enzo}, version 2.2 \citep{Enzo2013}.  A split hydro-solver with  the third order piece-wise parabolic (PPM) method was employed for solving the hydrodynamics, and the dark matter  was modeled using the particle-mesh technique. Self-gravity was calculated via a multigrid Poisson solver, and a network for the primordial chemistry was solved using the chemistry package \textsc{krome} \citep{Grassi2013} described in the next subsection. We followed the evolution of the halo starting from redshift z~=~99 with a top grid resolution of 128$^3$ cells. Two initial nested grids were subsequently added with a grid resolution of 128$^3$ cells each. A simulation box of cosmological size of 0.3 Mpc h$^{-1}$, centered on the most massive minihalo, was used. In total, we initialized 6291456 particles to compute the evolution of the dark-matter dynamics and obtained a final dark-matter resolution of 70 $\rm M_{\odot}$. The parameters for creating the initial conditions and the distribution of baryonic and dark-matter components were taken from the WMAP seven-year data \citep{Jarosik2011}. 

We furthermore allowed additional 27 levels of refinement in the central 18 kpc region of the halo during the course of simulation, which yielded a total effective resolution of 0.9 AU in comoving units. The resolution criteria used in these simulations are based on the Jeans length, the gas over-density, and the particle-mass resolution. We mandated a fixed Jeans resolution of 64 cells per Jeans length throughout the evolution of the simulations. It was suggested recently \citep{Federrath11,Turk2012,Latif13} that a resolution of at least 32 cells per Jeans length is needed to capture turbulent velocity fluctuations, for example. We stopped the simulations when they reached the highest refinement level.

\subsection{Chemistry}\label{sub:chem}
We followed the non-equilibrium evolution of nine species: H, H$^+$, H$^-$, H$_2$, H$_2^+$, He, He$^+$, He$^{2+}$, and e$^-$. A total of 21 kinetic reactions were included by using the rates discussed in \citet{Abel97} and \citet{Anninos97}. The heating and cooling rates  include H$_2$ formation heating as described in \citet{Omukai2000} , H$_2$ cooling as reported in \citet{GloverAbel08}, H and He collisional ionization, collisional excitation, recombination coolings, and bremsstrahlung cooling \citep{Cen92}. The heating due to the formation of H$_2$ from 3B processes was changed accordingly, following the rates of Table \ref{tab:three-body} to take into account the original CID rate from what the recombination was obtained by applying the detailed balance. 

The rate equations and the thermal evolution were solved by employing the new chemistry package \textsc{krome} \citep{Grassi2013}. Its implementation in \textsc{enzo} has previously been discussed in \cite{BovinoMNRASL}, where we showed the accuracy and stability of the \textsc{dlsodes} solver \citep{Hindmarsh83,Hindmarsh2005} included in the package. As we have shown there, the latter significantly improves the convergence of thermodynamical quantitites in high-resolution simulations. It is worth noting that we are considering an optically thin gas, which might lead to slightly lower temperatures.

\section{Results}\label{sec:results}
We performed a total of nine simulations to study the effect of the new H$_2$ 3B formation rates. For this purpose, we varied the 3B rates given in Table \ref{tab:three-body} and chose three different realizations of the cosmic density field, yielding minihalos with masses  of 1.3$\times$10$^5$~$\rm M_\odot$, 7$\times$10$^5$~$\rm M_\odot$, and 1$\times$10$^6$~$\rm M_\odot$, respectively. 

In Fig.s \ref{fig:chemA}, \ref{fig:chemB}, and \ref{fig:chemC} we report the averaged radial pofiles for the chemical species H, H$_2$, e$^-$, and H$^-$ of the nine different realizations. As stated in the introduction, we are interested in studying the effect of changing the H$_2$ 3B formation rate in our primordial chemical network and exploring different initial conditions.
 
The chemical abundances show similar general features for the three different halos employed: decreasing abundances of the electrons, H, and H$^-$ in favor of the formation of molecular hydrogen in the inner region ($R $$\le$ 10$^{15}$ cm) where, in fact, the H$_2$ profile shows a sudden increase. For different three-body rates, we found that the physical size of the molecular cores increases moving from a less efficient process (AB02) to the most efficient one (PSS83), which then has the largest molecular core, in agreement with the results of \citet{Turk2011ApJ}. This is also reflected in the atomic hydrogen  shown in the top right panels of the same figures, where clearly the H abundance significantly decreases at radii of 10$^{16}$ cm for the PSS83 case and slowly decreases for the less efficient rate reported by \citet{Abel2002}. The new rate calculated by \citet{Forrey2013} is, as expected, in between those of \citet{Abel2002} and \citet{Palla1983},  thus producing results for the H$_2$ abundance that agree with the overall trend. However, the details also depend on the halo under consideration, because it is closer to \citet{Abel2002} in halo B, closer to \citet{Palla1983} in halo C, and intermediate in halo A.

The dependence on the initial conditions is also visible in the electron abundance (bottom left panels) and in the H$^-$ profile (bottom right). These quantities do not necessarily lie in between the limiting cases of \citet{Palla1983} and \citet{Abel2002} either, as is visible in  Figure \ref{fig:chemA} for the electron mass fraction, where the new rates yield a lower electron mass fraction.
In Figs. \ref{fig:chemB} and \ref{fig:chemC}, on the other hand, the electron mass fraction is enhanced for radii $\le$ 10$^{15}$ cm in the F2013 case. 

The spherically averaged profiles for temperature, accretion rate, density, radial velocity, total energy, and Jeans mass are reported in Fig.s \ref{fig:figureA}, \ref{fig:figureB}, and \ref{fig:figureC} for the three different minihalos employed as a function of the radius and centered on the high-density peak. We only plot the region at which the core begins to become fully molecular and the 
new rates are relevant, corresponding to radii below 10$^{17}$ cm. On larger scales, we obtained results identical to those in \citet{BovinoMNRASL}. 

The thermal evolution (top right panels) is strongly affected by the molecular hydrogen evolution because it regulates the cooling, which thus depends on the changes in its formation rate. In Figure \ref{fig:figureA}, the only remarkable difference is in the AB02 run, where clearly the lower efficiency in forming H$_2$ via Eq. \ref{eq:formation} produces less cooling and leads to a higher temperature. The two runs produce a similar behavior with a temperature in the inner region around 400-500 K. This is consistent with the fact that the F2013 run approaches the fully molecular stage (i.e. $x_{H_2}\sim 1$) in a rather similar way to the simulations based on PSS83. For halo B in Figure \ref{fig:figureB}, on the other hand, the evolution is completely different. The AB02 run still produces the highest temperature ($\sim$1000 K) and F2013 follows a similar trend until a radius of 10$^{15}$ cm, when it slightly departs from the AB02 behavior. This is mainly because in F2013, the molecular stage is reached earlier. PSS83 again provides a very similar H$_2$ mass fraction. 

In Figure \ref{fig:figureC}, the results for halo C  are shown. In this case, we obtained the most surprising results: PSS83 and AB02 are still the two limiting cases because of their related 3B rates, but F2013 agrees very well with the AB02 temperature profile, which is not (at least at the first glance) strongly related to the H$_2$ behavior. The difference is also more pronounced if we compare the runs for  halo A (Fig. \ref{fig:figureA}) and  halo C (Fig. \ref{fig:figureC}), where a similar H$_2$ profile was obtained. In both cases, AB02 and F2013 have a similar molecular physical size, for instance, AB02 is fully molecular in both runs around 2$\times$10$^{14}$ cm and F2013 around 5-6$\times$10$^{15}$ cm, but the temperature profiles are very different. In halo A, AB02 is 200-300 K warmer, while in the halo C runs the temperatures are almost identical. We note in particular that the gas in halo A is initially colder with about 25 K at a radius of 5$\times$10$^{17}$ cm, while halos B and C have temperatures of $\sim$40 K at the same radial position. These differences naturally occur because of the different halo masses, collapse redshifts, and spin parameters.

The dynamical impact of these changes is explored in more detail in the same figures. In particular, we report relatively large differences in the radial velocities (middle right panels) where the magnitude changes by about 2-3 km/s and the position of the peak moves depending on the 3B rate employed, {which differs from the results reported by \citet{Turk2011ApJ} where the differences were up to 1 km/s and the peak position was quite consistent. This is more evident for halos B and C (and particularly pronounced in halo C), where the F2013 run produces a radial velocity four times higher than in the other runs. This causes an increase in the total energy (bottom left panels). Changes in the accretion rate (middle left panels), on the other hand, appear rather minor, and a similar result is obtained for the Jeans mass, which we computed as

\begin{equation}
	M_J = \left(\frac{3}{4\pi\rho}\right)^{1/2} \left(\frac{5 k T}{G\mu}\right)^{3/2},
\end{equation}
where $\rho$, $G$, and $T$ are the density, gravitational constant, and the temperature, respectively. The mean molecular weight $\mu$ is evaluated as
\begin{equation}
	\mu = \frac{m_px_{H} + 2 m_px_{H_2} + 4 m_p x_{He}}{x_{H_2} + x_H + x_{He}},
\end{equation}
where $m_p$ is the proton mass. Judging from these quantities, the resulting stellar masses may thus depend more on the initial conditions than the exact values of the three-body rates.
We note that a dependence of the final mass scale on the properties of the halo has also been reported by \citet{LatifBH,Latif2013arXiv} for atomic cooling halos.

To explore the structure in the collapsing halos, Figure \ref{fig:projection} shows a comparison of the morphology for AB02, PSS83, and F2013. The density projections for the three minihalos reflect the results reported and discussed in Figs. \ref{fig:figureA}, \ref{fig:figureB}, and \ref{fig:figureC} for the thermal evolution, infall velocity, and Jeans mass. In addition, there are  relevant differences in the morphology, because halo A (MHA) is almost spherically symmetric while halos B (MHB) and C (MHC) show a more turbulent structure. Comparing the the same halo for the three different rates employed here, the differences in morphology seem to depend more on the halo than the three-body rate: (i) halo A is largely  spheroidal/ellipsoidal in all the cases and the only difference lies on the more massive central core obtained by employing the AB02 rate. This is caused by the higher temperature reached at that stage compared with the PSS83 and F2013 realizations; (ii) halo B seems to go from a turbulent large structure to a more compact one when going from AB02 to F2013. Owing to the lower temperature achieved by the PSS83 and F2013, we note here some turbulent-clumpy structures in the center of the collapse, which are not present in the AB02 run. In these realizations, we also see a more compact structure when we use the F2013 rate; (iii) finally, halo C seems to exhibit either a spherical (AB02), turbulent (F2013), or disk-like structure (PSS83) depending on the three-body rate. It is worth noting that we can only assess general and qualitative statements about the morphology of the collapsing cloud and that a deeper analysis and longer evolution are needed to explore possible fragmentation processes.

To conclude our analysis, we show in Figure \ref{fig:projection2} the vorticity squared, the temperature, and the H$_2$ projections for the realizations performed by using the new accurate quantum-mechanical H$_2$ 3B formation rate \citep{Forrey2013}. These quantities reflect the behavior discussed above and show how much the initial conditions can affect the morphology of the collapsing cloud. While we recommend now using the correct three-body rates provided by \citet{Forrey2013}, it appears at least equally important to explore larger samples to capture the range of possible initial conditions.


    \begin{figure*}
   \centering
   \includegraphics[width=0.8\textwidth]{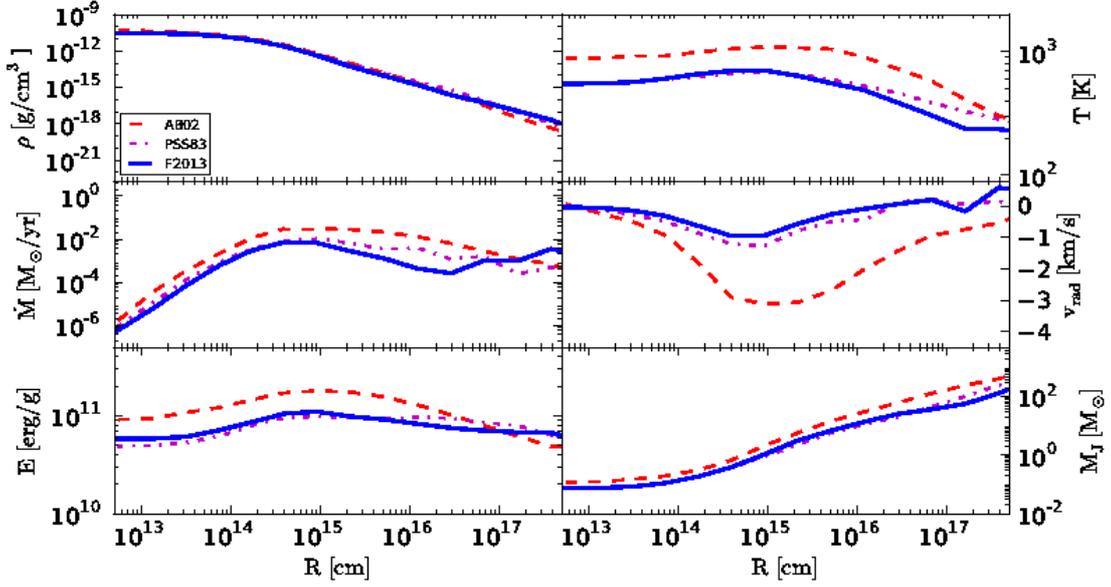}
      \caption{Radially spherically averaged profile of H, H$_2$, H$^-$, and electron mass fractions as a function of the radius, taken at the densest point in the calculations. The results are shown for the three different H$_2$ 3B formation rates reported in Table \ref{tab:three-body} and for halo A. }
         \label{fig:chemA}
   \end{figure*}

 \begin{figure*}
   \centering
   \includegraphics[width=0.8\textwidth]{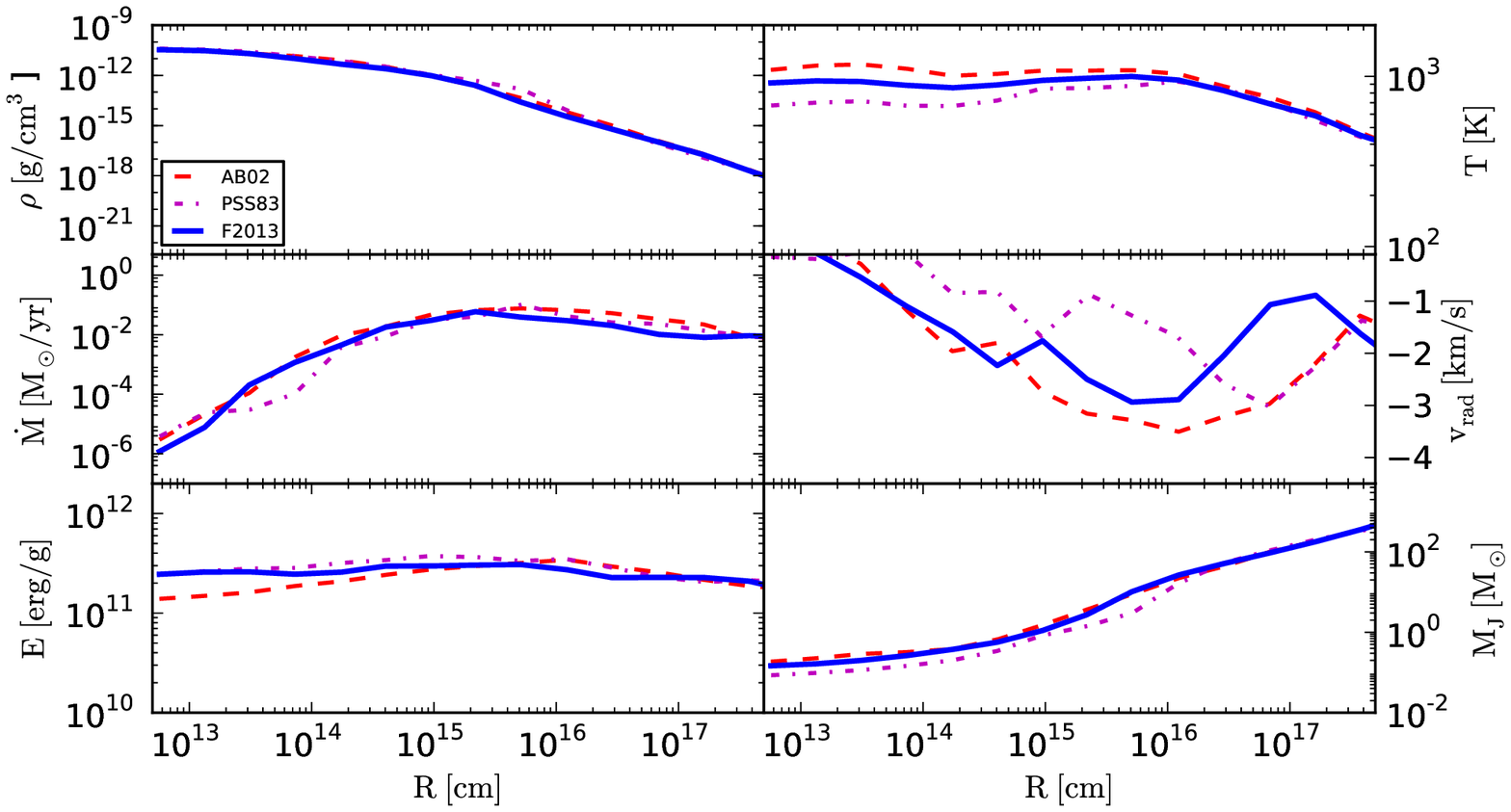}
      \caption{Radially spherically averaged profile of H, H$_2$, H$^-$, and electron mass fractions as a function of the radius, taken at the densest point in the calculations. The results are shown for the three different H$_2$ 3B formation rates reported in Table \ref{tab:three-body} and for halo B. }
         \label{fig:chemB}
   \end{figure*}
   
    \begin{figure*}
   \centering
   \includegraphics[width=0.8\textwidth]{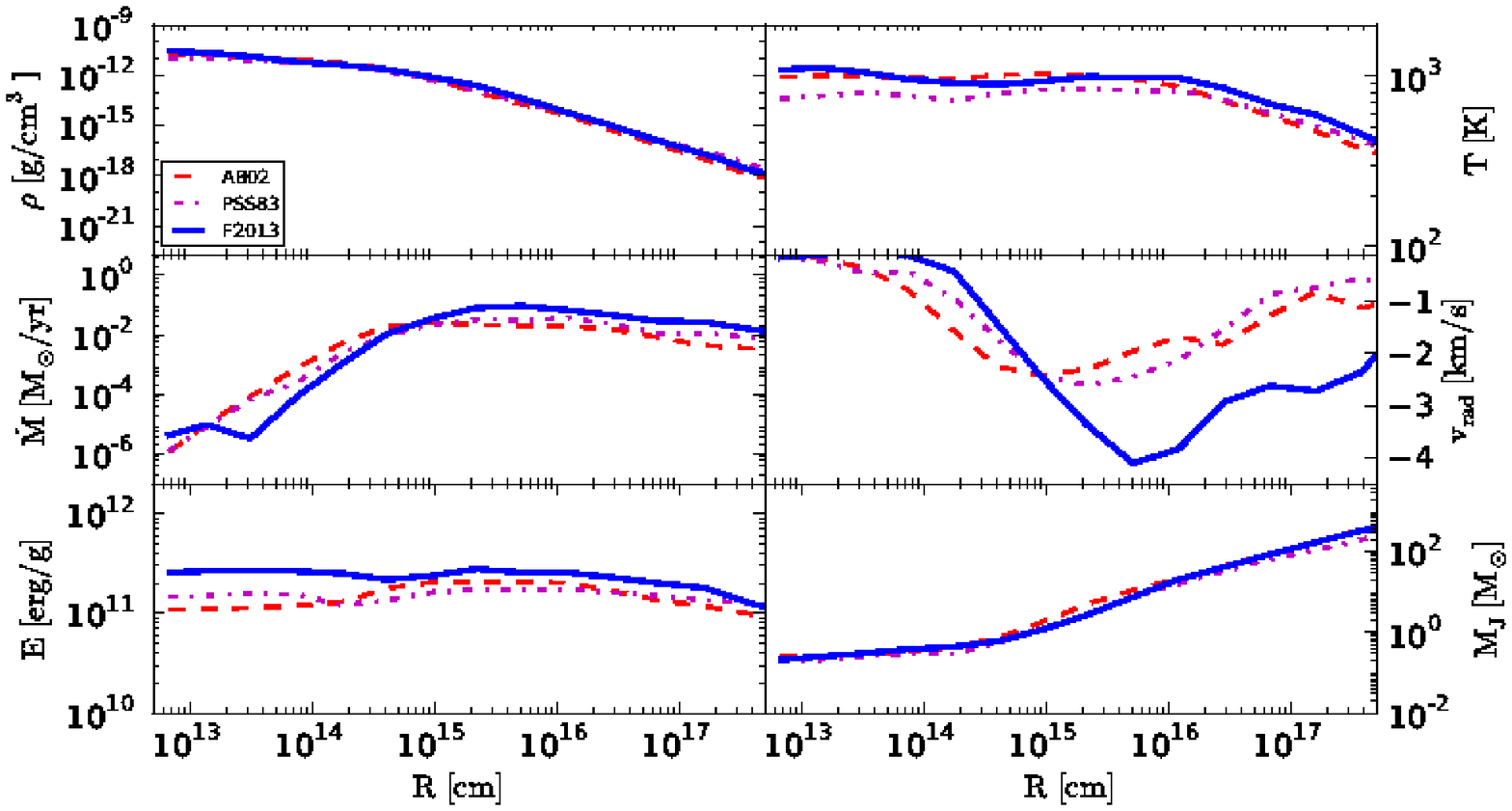}
      \caption{Radially spherically averaged profile of H, H$_2$, H$^-$, and electron mass fractions as a function of the radius, taken at the densest point in the calculations. The results are shown for the three different H$_2$ 3B formation rates reported in Table \ref{tab:three-body} and for halo C. }
         \label{fig:chemC}
   \end{figure*}

 \begin{figure*}
   \centering
   \includegraphics[width=0.8\textwidth]{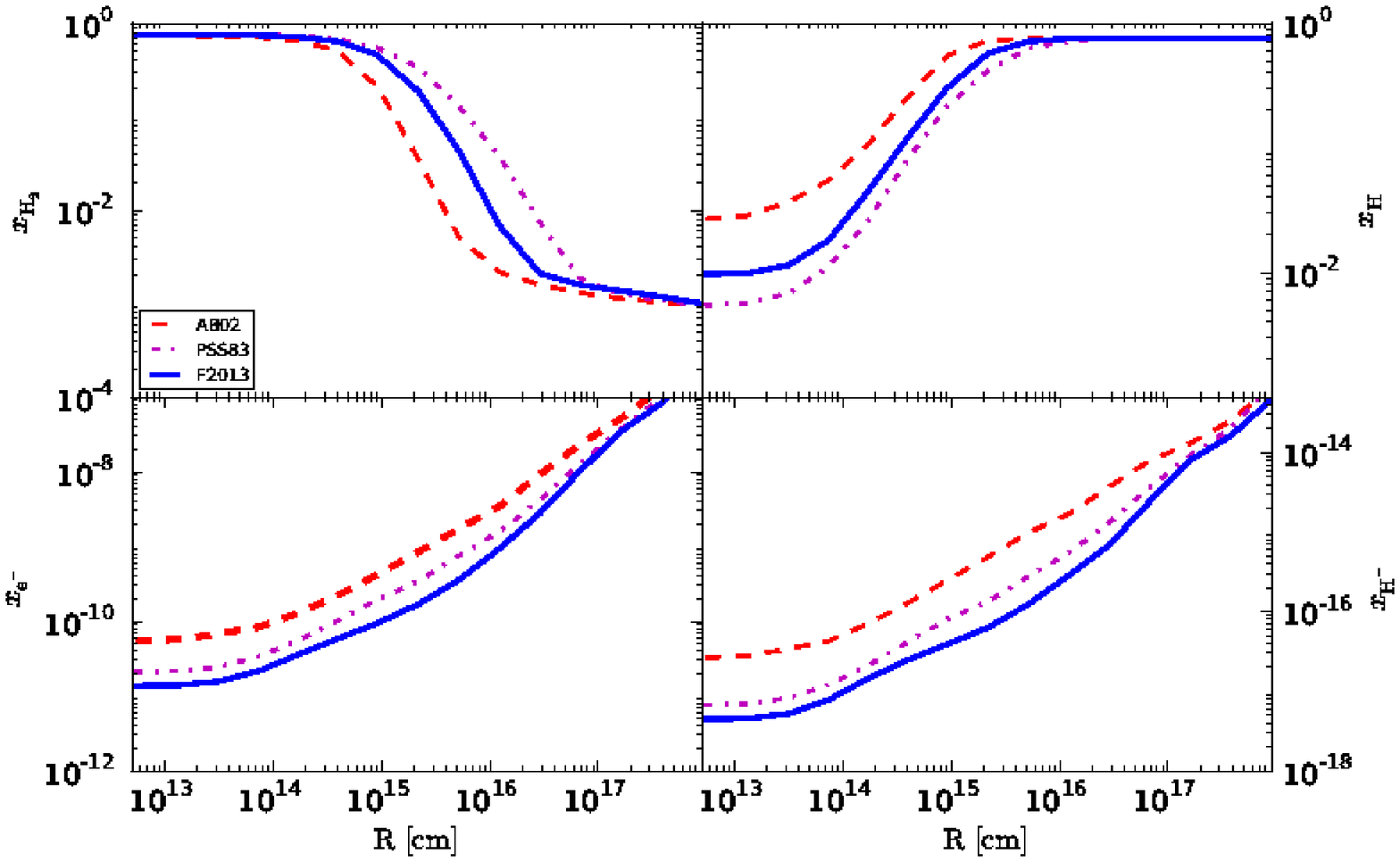}
      \caption{Radially spherically averaged profile of density (upper left), temperature (upper right), accretion rate (middle left), infall velocity (middle right), total energy (bottom left), and Jeans mass (bottom right) for halo A and the three different H$_2$ 3B formation rates reported in Table \ref{tab:three-body}. }
         \label{fig:figureA}
   \end{figure*}

 \begin{figure*}
   \centering
   \includegraphics[width=0.8\textwidth]{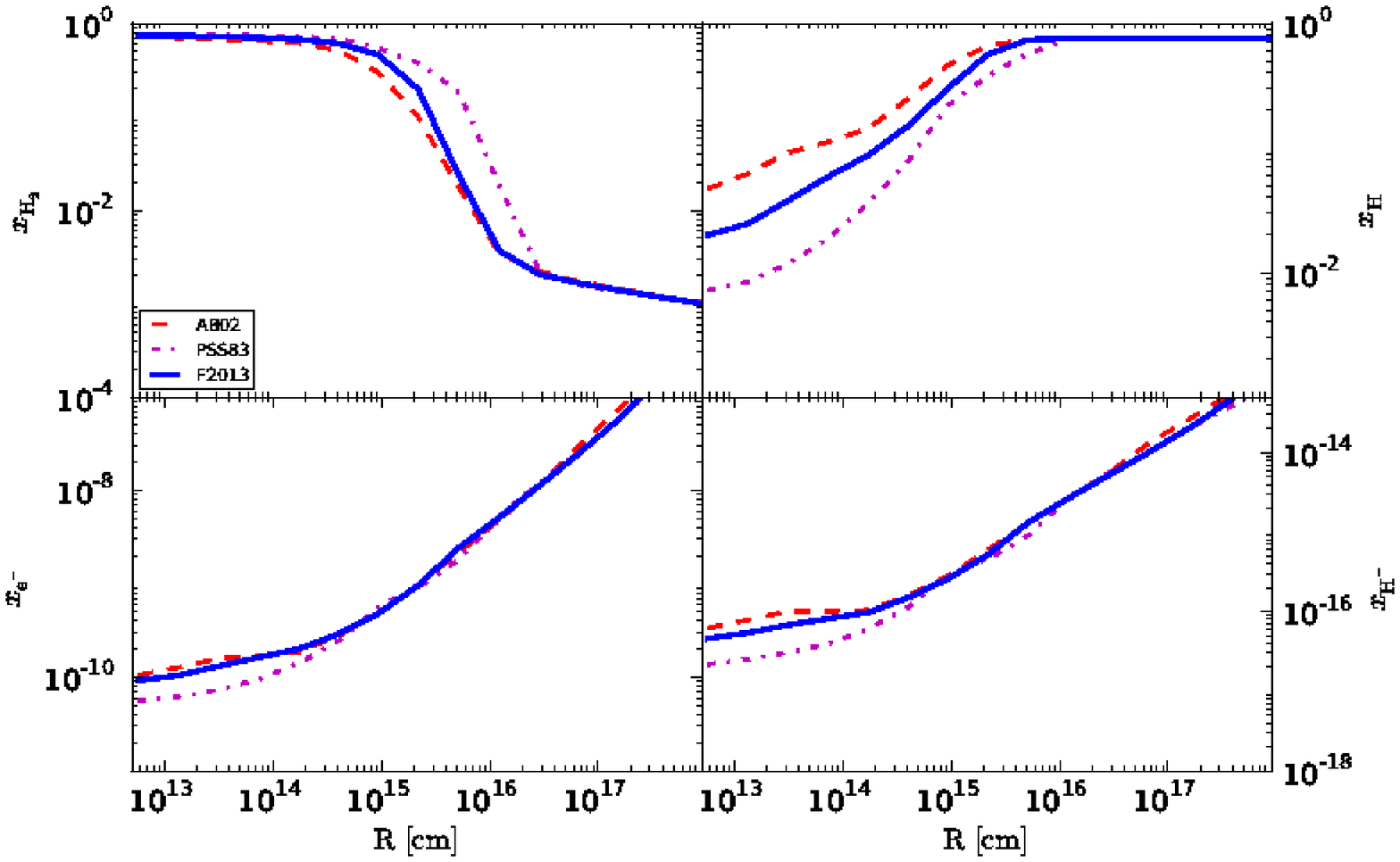}
      \caption{Radially spherically averaged profile of density (upper left), temperature (upper right), accretion rate (middle left), infall velocity (middle right), total energy (bottom left), and Jeans mass (bottom right) for halo B and the three different H$_2$ 3B formation rates reported in Table \ref{tab:three-body}.}
         \label{fig:figureB}
   \end{figure*}
   
    \begin{figure*}
   \centering
   \includegraphics[width=0.8\textwidth]{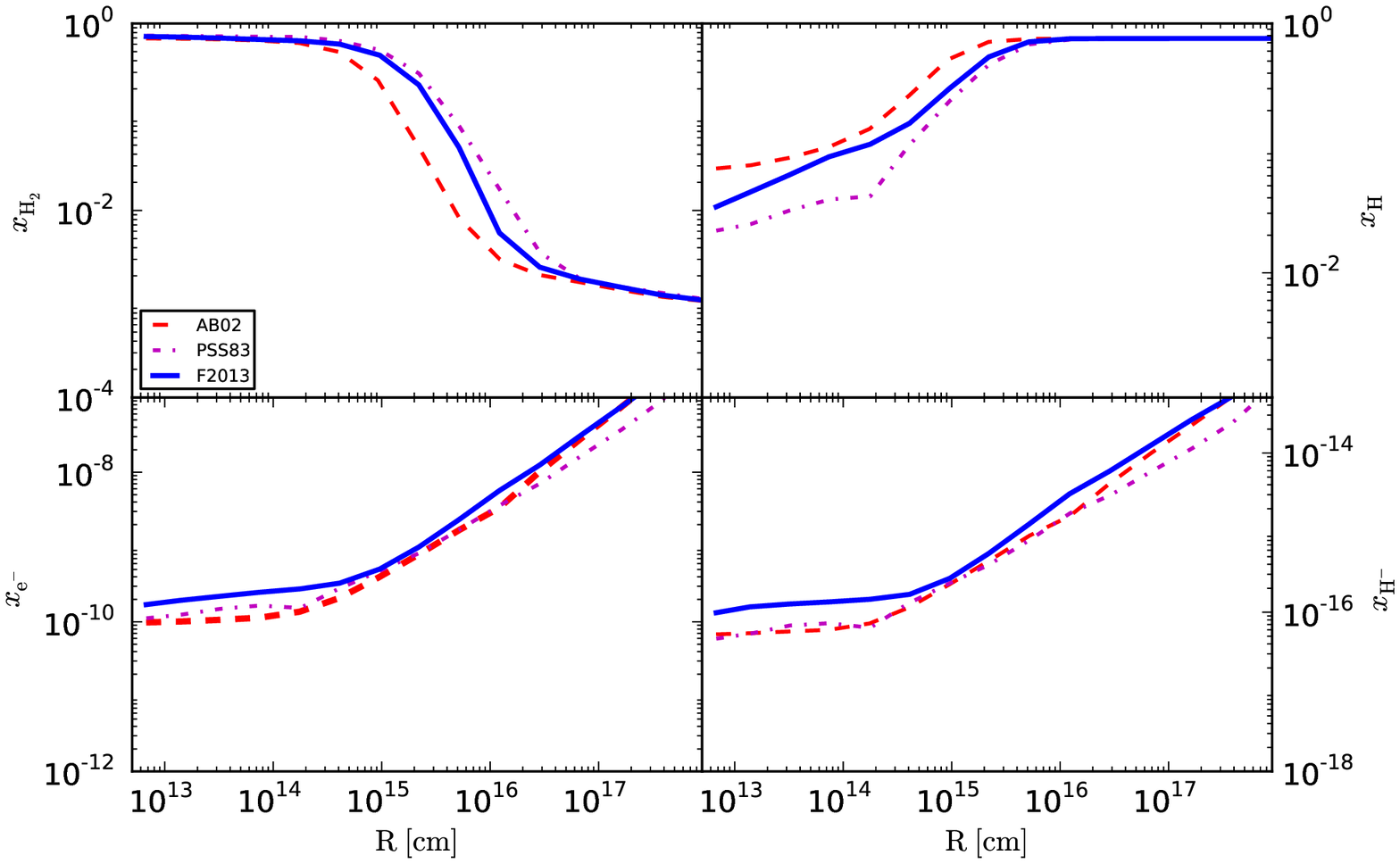}
      \caption{Radially spherically averaged profile of density (upper left), temperature (upper right), accretion rate (middle left), infall velocity (middle right), total energy (bottom left), and Jeans mass (bottom right) for halo C and the three different H$_2$ 3B formation rates reported in Table \ref{tab:three-body}.}
         \label{fig:figureC}
   \end{figure*}

    \begin{figure*}
   \centering
   \includegraphics[width=0.8\textwidth]{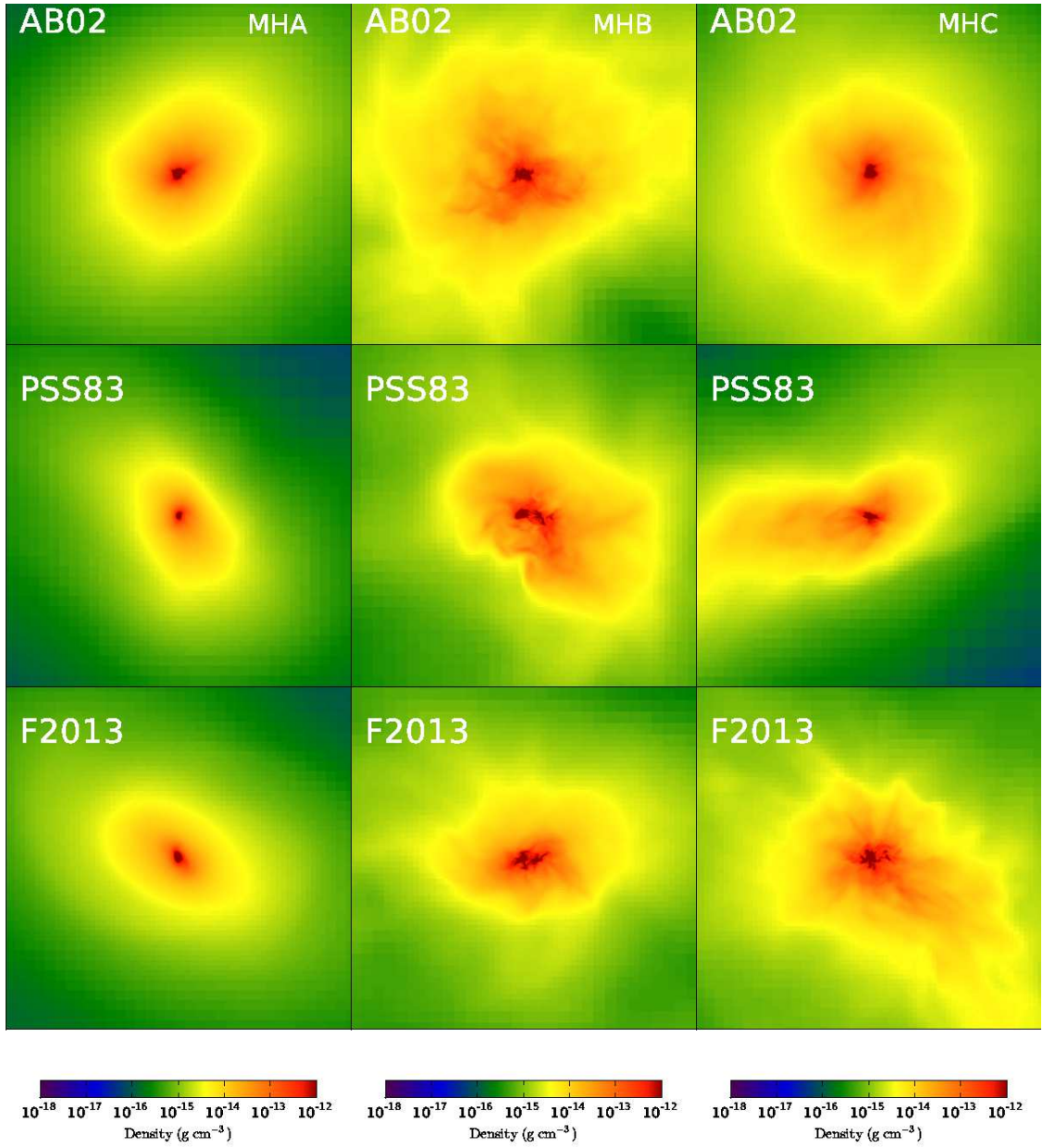}
      \caption{Density projection for the realizations. From the top: AB02, PSS83, and F2013 runs and from left to right halo A (MHA), halo B (MHB), and halo C (MHC), respectively.}
         \label{fig:projection}
   \end{figure*}

  \begin{figure*}
   \centering
   \includegraphics[width=0.8\textwidth]{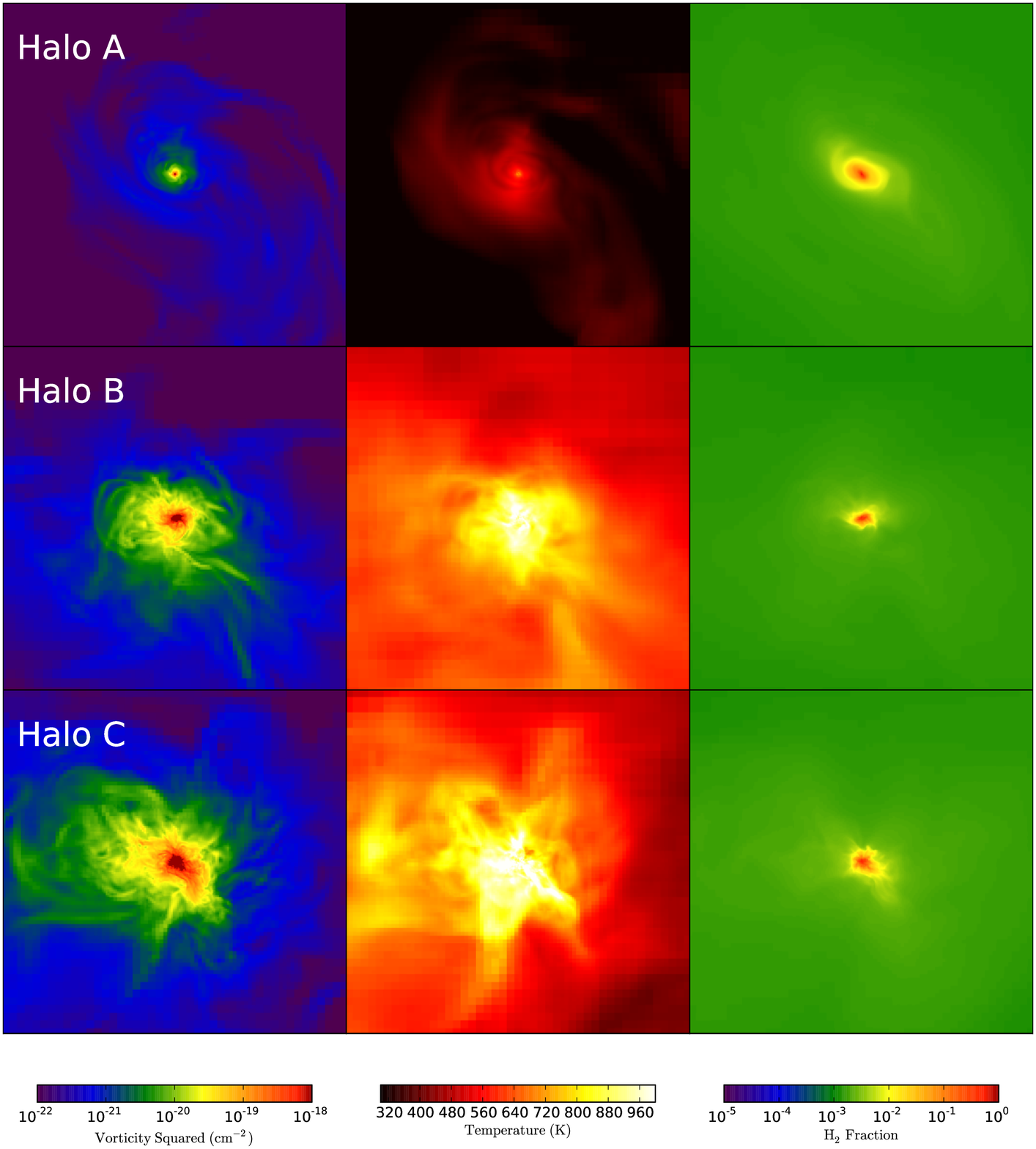}
      \caption{Vorticity squared, temperature, and H$_2$ mass fraction, and x-axis projections for the three different halos employed in the calculations (from the top to the bottom: halo A, halo B, and halo C) and for the new accurate 3B formation rate of Eq. \ref{eq:formation}.}
         \label{fig:projection2}
   \end{figure*}

\section{Conclusions}\label{sect:conclusions}
We studied the effect of the  three-body H$_2$ formation rate in Population III star formation simulations, employing the recent accurate quantum-mechanical calculation by \citet{Forrey2013}. We compared the relative importance of correct three-body rates with the potential effect of different initial conditions. 

We performed nine simulations for three different minihalos and three different 3B formation rates commonly used in the literature, including those by \citet{Palla1983}, \citet{Abel2002}, and the new rates by \citet{Forrey2013}, and analyzed their impact on chemistry, thermal evolution, and the dynamics of the collapse. Our simulations started from  cosmological initial conditions and employed the adaptive mesh refinement technique until we reached the highest refinement level at densities around 10$^{-11}$ g cm$^{-3}$.

While one might naively expect  the  new \citet{Forrey2013} rates to lead to a behavior in between the simulations based on \citet{Palla1983} and \citet{Abel2002}, the actual evolution is sometimes closer to \citet{Abel2002}, and sometimes closer to \citet{Palla1983}, depending on the initial conditions. We note that situations may even occur where the H$_2$ abundance evolves according to the \citet{Abel2002} runs, while the thermodynamics are closer to the run based on \citet{Palla1983}, and vice versa. The latter shows that a different H$_2$ formation rate gives rise to a different nonlinear evolution, which thus has nontrivial consequences on the results. To understand the behavior of a particular halo, it is thus mandatory to employ accurate rates. 

While the accretion rates and the Jeans mass only show a minor dependence on the reaction rates and a stronger dependence on the initial conditions, an accurate modeling is nevertheless relevant. In addition, it shows now an increasing need to study larger samples to understand the expected statistical distribution in halo properties. In addition, it will be important to assess the impact on fragmentation by continuing these simulations beyond the formation of the first peak and following the accretion onto the central clump and also the potential formation of additional clumps. 

We note that during the collapse, radiative feedback from the protostar will become relevant and may limit the subsequent accretion. Studies by \citet{Hosokawa11} and \citet{Susa2013} indeed suggested a characteristic mass scale of the order of 50 M${_\odot}$. \citet{Hirano2013arXiv} investigated a total of 100 minihalos, confirming these results, but showing a considerable distribution of the stellar masses around the mean. This confirms our assessment that the initial conditions have a strong impact on the final results. We also note that magnetic fields are efficiently amplified during the collapse in minihalos \citep{Schleicher2010,Sur10,Schober2012,Turk2012,Bovino2013NJP} and can subsequently suppress fragmentation and change the final mass scale \citep{Machida2008, Machida2013arXiv}. A treatment of these effects will therefore be necessary in future studies.

The results presented here thus confirm the potential importance of choosing the correct reaction rates, but also hint at the potentially larger relevance of the initial conditions. This is in line with previous results obtained by \citet{Jappsen2009}, showing that the initial conditions can be more relevant than metallicity thresholds. While this also makes sense for a primordial gas, it is still reassuring that an important uncertainty has been removed via quantum-chemical calculations.
 
\section*{Acknowledgements}
S.B. thanks for funding through the DFG priority programme `The Physics of the Interstellar Medium' (project SCHL 1964/1-1). D.R.G.S. thanks for funding via the SFB 963/1 on "Astrophysical Flow Instabilities and Turbulence" (project A12). T.G. thanks the CINECA consortium of the awarding of financial support while the present research was carried out. The simulation results are analyzed using the visualization toolkit for astrophysical data YT \citep{Turk2011yt}.

\bibliographystyle{mn2e}      
\bibliography{mybib_new}

\end{document}